\documentclass[10pt,conference]{IEEEtran}

\usepackage{graphicx}   

\usepackage{subfigure} 

\usepackage{url}        

\usepackage{amsmath}    
\usepackage[all]{xy}
\usepackage{amssymb}
\usepackage{caption2}
\usepackage{comment}
\usepackage{mathbbol}

\newtheorem{theo}{Theorem}
\newtheorem{prop}[theo]{Proposition}

\newtheorem{defi}[theo]{Definition}

\newcommand{\C}{{\cal C}}
\renewcommand{\O}{{\cal O}}
\renewcommand{\L}{{\cal L}}
\newcommand{\F}{{\mathbb F}}
\newcommand{\Div}{{\mbox{\rm Div}}}
\newcommand{\ev}{{\mbox{\rm ev}}}

\renewcommand{\k}{\mathbb{k}}

\begin{document}
%
\title{Algebraic-geometric codes from vector bundles\\ and their decoding}
%
\author{Valentin~Savin, CEA-LETI, MINATEC, Grenoble, France, valentin.savin@cea.fr
\thanks{Manuscript received {\it Month Day, Year}; revised {\it Month Day, Year}.}
}
%
%
%
%
\markboth{IEEE Transactions on Information Theory, to be submitted}{Shell \MakeLowercase{\textit{et al.}}:
Bare Demo of IEEEtran.cls for Journals}
%



\maketitle

\begin{abstract}
Algebraic-geometric codes can be constructed by evaluating a certain set of functions on a set of distinct rational points of an algebraic
curve. The set of functions that are evaluated is the linear space of a given divisor or, equivalently, the set of section of a given line
bundle. Using arbitrary rank vector bundles on algebraic curves, we propose a natural generalization of the above construction. Our codes
can also be seen as interleaved versions of classical algebraic-geometric codes. We show that the algorithm of Brown, Minder and
Shokrollahi can be extended to this new class of codes and it corrects any number of errors up to $t^{*} - g/2$, where $t^{*}$ is the
designed correction capacity of the code and $g$ is the curve genus.
\end{abstract}

\begin{keywords}
Algebraic-geometric codes, vector bundles.
\end{keywords}

%
\IEEEpeerreviewmaketitle

\section{Introduction}

The construction of error correcting codes using methods from algebraic-geometry was first proposed by Goppa \cite{goppa_curves},
\cite{goppa_ag_codes} in the early '80s. He constructed codes by evaluating a certain set of differential forms on a set of distinct
rational points of an algebraic curve. From a dual point of view, Goppa's codes can be constructed by evaluating a certain set of functions
on a set of distinct rational points of an algebraic curve \cite{ag_codes_survey}. The set of functions that are evaluated is the linear
space of a given divisor, whose support is disjoint from the set of evaluation points. These codes generalize the
Bose-Chaudhuri-Hocquenghem (BCH), Reed-Solomon (RS) and Goppa codes (latest codes being introduced by Goppa in the early '70s). Unlike the
RS codes, algebraic-geometric (AG) codes are not generally MDS codes, but their Singleton defect is upper bounded by the genus of the
curve. However, despite their Singleton defect, AG codes are better than RS codes since they allow the construction of longer codes over
the same alphabet. Another advantage of AG codes is that for fixed code parameters, their encoding and decoding algorithms run faster as
they can be performed in a smaller field. Since the end of the 80s, intensive research has been done on decoding algorithms and most of the
methods used to decode BCH, RS or Goppa codes were extended to the class of AG codes \cite{decoding_ag_survey}.

In order to explain our approach to construct codes from vector bundles on projective curves, let us recall the classical construction of
AG codes. Let $\C$ be an absolutely irreducible, smooth, projective curve of genus $g$, defined over a finite base field $\F_q$. Let ${\cal
P} = \{P_1, P_2, \dots, P_n\}$ be a set of distinct rational points of $\C$ and $D\in \Div(\C)$ be a divisor whose support is disjoint from
the set ${\cal P}$. The linear code $C({\cal P},D)$ is defined as the image of the evaluation map:
\begin{equation}\label{constr_ag_codes}
\begin{array}{c@{\,\,}c@{\,\,}ccc}
\ev & : & \L(D) & \longrightarrow & \F_q^n  \\
    &   &  f    & \longmapsto     & (f(P_1), f(P_2), \dots, f(P_n))
\end{array}
\end{equation} where $\L(D)$ is the linear space of $D$.
The parameters of the code, or bound on them, can be determined using well-known statements in algebraic-geometry, notably the Hasse-Weil
theorem and the Riemann-Roch theorem, and it can be seen that the Singleton defect of the code is upper bounded by the curve genus $g$.

Interleaved AG codes were defined in \cite{interleaved_ag} as follows. Suppose that $Q=q^r$ and identify $\F_Q$ and $\F_q^r$ as
$\F_q$-vector spaces by fixing a basis of $\F_Q$ over $\F_q$. For any point $P\in\C$ and any $f=(f_1, f_2, \dots, f_r)\in\L(D)^{\oplus r}$,
the evaluation vector $f(P) = (f_1(P), f_2(P), \dots, f_r(P))\in \F_q^r$ can be identified with an element of $\F_Q$. The code $C({\cal
P},D,r)$ is defined as the image of the evaluation map
\begin{equation}\label{constr_inter_ag_codes}
\begin{array}{c@{\,\,}c@{\,\,}ccc}
\ev & : & \L(D)^{\oplus r} & \longrightarrow & \F_Q^n  \\
    &   & f    & \longmapsto     & (f(P_1), f(P_2), \dots, f(P_n))
\end{array}
\end{equation}
This code does not generally be $\F_Q$ linear, however it is a $\F_q$-vector subspace of $\F_Q^n$.

To explain our approach, we first remark that there is a one-to-one correspondence between the divisor class group $\mbox{Cl}(\C)$ (the
group of equivalence classes of $\Div(\C)$) and isomorphism classes of line bundles on $\C$. For any divisor $D$, we denote by $\O(D)$ the
line bundle associated with $D$. The linear space $\L(D)$ is isomorphic to the space of global sections of $\O(D)$, which is generally
denoted by $H^0(\C, \O(D))$.

We construct AG codes by evaluating global sections of arbitrary rank vector bundles $E$ on the points $P_1, P_2, \dots, P_n$. Thus, the
classical construction of AG codes (\ref{constr_ag_codes}) corresponds to the case $E = \O(D)$ and the construction of interleaved AG codes
(\ref{constr_inter_ag_codes}) corresponds to $E = \O(D)^{\oplus r}$. In general, if $E$ is a rank-$r$ vector bundle, we obtain a code
$C({\cal P},E)$ over $\F_Q$, not necessarily linear, where $Q=q^r$. Further, we investigate the parameters of these codes, or bound on
them, such as the code length, dimension, and minimum distance. While the code dimension can still be lower bounded by the Riemann-Roch
theorem, it is much harder to compute its exact value or to give a lower bound on the minimum distance. The main reason is that unlike line
vector bundles, arbitrary rank vector bundles of negative degree may have non zero sections. To overcome such situations, we need the
vector bundle to satisfy some stability condition. When this condition is satisfied, we can compute the code dimension and we can show that
the Singleton defect is upper bounded by the curve genus $g$.

Now we come to the decoding problem. Bleichenbacher, Kiayias and Yung \cite{interleaved_rs} proposed a new decoding algorithm for
interleaved Reed-Solomon codes over the $Q$-ary symmetric channel, which was later extended by Brown, Minder and Shokrollahi to the case of
interleaved AG codes \cite{interleaved_ag}. One advantage of using interleaved AG codes is that they allow transmissions at rates closer to
the channel capacity. If $Q=q^r=2^{hr}$, the $Q$-ary symmetric channel model applies to settings where packets of $hr$ bits are sent and
errors are assumed to be bursty. From the coding theory perspective, errors on bits of the same packet are assumed to be correlated. This
usually arrives when packets of bits are sent over different transmission channels, some of which may induce errors. When $Q$ is too large,
efficient decoding of codes designed over $\F_Q$ is impossible, which explains the advantage of interleaved codes, since their decoding
algorithms operate over $\F_q$. In this paper we show that the decoding algorithm of interleaved AG codes can be extended to the class of
 codes constructed from vector bundles and it corrects any number of errors up to $g/2$ from half the designed minimum distance of the
code.

In the next section we review some of the mathematical background needed to understand the construction of AG codes from vector bundles on
algebraic curves. Our aim is not to do an exhaustive nor a self-contained presentation, but rather to guide the reader from elementary to
more complicated objects and show that objects as vector bundles or stable vector bundles naturally arise in algebraic geometry. In
sections \ref{sec:ag_codes} and \ref{sec:decoding} we present respectively the construction of AG codes from vector bundles and their
decoding over the $Q$-ary symmetric channel. Section \ref{sec:construction_vb} focuses on the construction of vector bundles verifying the
stability condition. Finally, section \ref{sec:conclusions} concludes this paper.

\section{Vector bundles on projective curves}
For more details on this topic we refer to classical algebraic-geometry texts, such as \cite{hartshorne}. We assume that the reader is
familiar with classical algebraic-geometric codes and basics on algebraic-geometry, such as curves and divisors.

We first deal with algebraic closed fields, then we generalize to arbitrary fields by extending scalars to their algebraic closure. Let
$\C$ be an irreducible, smooth, projective curve of genus $g$, over an algebraic closed field $\k$ (although these conditions are not
always necessary). There are two possibilities of introducing vector bundles over $\k$: they can be defined as locally free sheaves of
finite rank over the curve $\C$, or as families of vector spaces parameterized by $\C$. While the first definition is often more convenient
for deeper analysis and understanding, the second definition has the advantage of being more intuitive and comprehensible to those not
familiar with heavy algebraic-geometry formalism. We will focus on intuition and will introduce vector bundles as families of vector spaces
parameterized by $\C$. Precisely, a {\em rank-$r$ vector bundle} over $\C$ is a variety $E$ together with a surjective regular map
$E\stackrel{\pi}{\rightarrow}\C$, such that:
\begin{enumerate}
\item for any $P\in\C$ the fiber $E_P := \pi^{-1}(P)$ is endowed with a structure of $\k$-vector space
of dimension $r$,
\item for any $P\in\C$, there is an open subset $U\in\C$ containing $P$ and a map $\varphi : U
\times \k^r \rightarrow \pi^{-1}(U)$ such that:\\
 -- $\varphi(P', v) \in E_{P'}$ for any $P'\in U$\\
 -- the restriction $\varphi : \{P'\} \times \k^r \rightarrow E_{P'}$ is a vector bundles isomorphism for any $P'\in U$
\end{enumerate}
It may be convenient to visualize the second condition using the following commutative diagram:
$$\xymatrix@C=10pt@R=18pt{%
 U \times \F_q^r \ar@{->}[rr]^(.4){\varphi}\ar@{->}[dr]_{\mbox{pr}_U} & &  \pi^{-1}(U) =: E_U \ar@{->}[dl]^{\pi}\\
 & U}$$
One may think of a vector bundle as a family of vector spaces $\{E_P\}_{P\in \C}$ parameterized by $\C$, which looks locally trivial. Any
algebraic operation with vector spaces can be extended to vector bundles: for instance, we can define direct sums, tensor products,
exterior (wedge) products and dual vector bundles.

A {\em (global) section} of the vector bundle $E$ over $\C$ is a regular map $s:\C\rightarrow E$, such that $s(P)\in E_P$ for any $P\in
\C$. The set of global sections of $E$ is denoted by $H^0(\C, E)$. It is canonically endowed with a $\k$-vector space structure and its
dimension is denoted by $h^0(\C, E)$.

A {\em line bundle} $L$ over $\C$ is simply a rank-$1$ vector bundle. To any meromorphic section $s:\C\rightarrow L$ we associate a divisor
$(s) = Z(s)-P(s)$, where $Z(s)$ and $P(s)$ denote respectively the set of zeros and the set of poles, counted with multiplicities. Note
that if $s\in H^0(\C, L)$ is a global section, then $(s) = Z(s)$ is an effective divisor. The {\em degree} of $L$ is by definition the
degree of $(s)$.
$$\deg(L) := \deg(s)$$
The fact that $\deg(L)$ is well defined follows from the first assertion of the following proposition. The second assertion highlights the
connection between global sections of line bundles and linear spaces.
\begin{prop}
$(a)$ If $s$, $s'$ are meromorphic sections of a line bundle $L$, then $(s)$ and $(s')$ are linear equivalent
divisors.

$(b)$ For any meromorphic section $s$, the map
$$\begin{array}{ccc}
\L(s) & \longrightarrow & \ H^0(\C, L)\\
 f    & \longmapsto     & fs
 \end{array}$$
 defines an isomorphism of vector spaces.
\end{prop}

If $E$ is a rank-$r$ vector bundle on $\C$, its $r$-th exterior power $\det(E) := \wedge^rE$ is a line bundle, called the {\em determinant
bundle} of $E$. The {\em degree} of $E$ is by definition the degree of its determinant bundle:
$$\deg(E) := \deg(\det(E))$$
The {\em slope} of $E$ is defined by $\mu(E)= \displaystyle\frac{\deg(E)}{\mbox{\rm rank}(E)}$.

Examples of vector bundles that naturally arise in algebraic-geometry are the tangent bundle and its dual, called the cotangent bundle, or
if the curve $\C$ is embedded in some projective space, the normal bundle. The cotangent bundle of $\C$ is also called the {\em canonical
bundle} and is denoted by $\Omega$. Any divisor associated with the canonical bundle is called {\em canonical divisor}. We can now state
the Riemann-Roch theorem \cite{hartshorne}.

\begin{theo}[Riemann-Roch]
Let $E$ be a vector bundle of rank $r$  and degree $e$ on $\C$. Then:
$$h^0(\C,E) - h^0(\C, \Omega\otimes E^{*}) = e + r(1-g)$$
where $E^{*}$ is the dual vector bundle of $E$.
\end{theo}

At this point we have introduced the necessary tools for defining AG codes from vector bundles. In order to be able to investigate the
parameters of these codes and their decoding algorithm, we need the vector bundles to satisfy some stability condition. The theory of
stable vector bundles goes back to the classification problem of vector bundles in the 60s. However, in this paper we need only a weaker
version of the stability, which we call weak stability. Before introducing the stability condition, we have to be more specific about
morphisms of vector bundles and vector sub-bundles.

Let $E$ and $F$ be two vector bundles on $\C$. A {\em morphism of vector bundles} is a regular map
$\varphi:E\rightarrow F$, such that:
\begin{itemize}
\item for any $P\in\C$ and any $x\in E_P$, $\varphi(x)\in F_P$,
\item for any $P\in\C$ the induced map $\varphi_P:E_P\rightarrow F_P$ is a morphism of vector spaces.
\end{itemize}
Any morphism of vector bundles $\varphi:E\rightarrow F$  induces in a obvious way a morphism between the corresponding vector spaces of
global sections, that is $\varphi:H^0(C,E) \rightarrow H^0(C, F)$.

 We say that $E$
is a {\em sub-bundle} of $F$ if there is a morphism of vector bundles $\varphi:E\hookrightarrow F$, such that for any point $P\in\C$ the
induced morphism $\varphi_P:E_P\rightarrow F_P$ is injective. In this case one can define a {\em quotient vector bundle} $F/E$, whose fiber
in a point $P\in\C$ is defined by $(E/F)_P = E_P/F_P$.

\begin{defi}
A vector bundle $E$ is said to be {\em weakly stable} if for any line sub-bundle $L\subset E$ the following inequality holds:
$$\deg(L) \leq \mu(E)$$
\end{defi}

\begin{prop}
$(a)$ Any line bundle is weakly stable.

\noindent $(b)$ Let $E=\displaystyle \mathop{\oplus}_{i=1}^{n} E_i$ be a direct sum of vector bundles. Then $E$ is weakly stable if and
only if $\mu(E_1) =  \dots = \mu(E_n)$ and all $E_i$, $i=1,\dots, n$, are weakly stable.

\noindent $(c)$ If $E$ is weakly stable then for any line bundle $L$, the tensor product $E\otimes L$ is also weakly stable.
\end{prop}

\begin{prop}
Assume that $E$ is a weakly stable vector bundle on $\C$.

\noindent $(a)$ Any global section of $E$ vanishes in at most $\lfloor\mu(E)\rfloor$ points.

\noindent $(b)$ If $\deg(E) < 0$ then $h^0(\C, E) = 0$.
\end{prop}

\section{AG codes from vector bundles}\label{sec:ag_codes}

In this section we define algebraic-geometric codes from vector bundles. Through the rest of this paper, we denote by $\C$ an absolutely
irreducible, smooth, projective curve of genus $g$, defined over a finite base field $\F_q$. For any vector bundle $E\rightarrow \C$, let
$\bar E\rightarrow \bar\C$  be the vector bundle obtained by extending scalars from $\F_q$ to its algebraic closure $\bar\F_q$. We define
$\deg(E) = \deg(\bar E)$ and we say that $E$ is weakly stable iff $\bar E$ is weakly stable.

Let ${\cal P} = \{P_1,\dots,P_n\}$ be a set of distinct rational points of $\C$, $E$ be a rank-$r$ vector bundle on $\C$, and set $Q =
q^r$.
We fix once for all basis of $\F_Q$ and $E_{P_i}$,\break $i=1,\dots,n$, as vector spaces over $\F_q$, which allows us to identify
$E_{P_i}\simeq \F_Q \simeq \F_q^r$. Henceforth, these identifications will be used without recalling the subjacent basis. The
algebraic-geometric code $C({\cal P},E)$ over $\F_Q$ is defined as the image of the evaluation map:
\begin{equation}\label{constr_ag_codes_vb}
\begin{array}{c@{\,\,}c@{\,\,}ccc}
\ev & : & H^0(\C, E) & \longrightarrow & \displaystyle\mathop{\oplus}_{i=1}^n E_{P_i} \simeq \F_Q^n  \\ 
    &   & f    & \longmapsto     & (f(P_1), f(P_2), \dots, f(P_n))
\end{array}
\end{equation}
Note that this is not necessarily a $\F_Q$ linear code, but it is a $\F_q$ linear subspace of $\F_Q^n$. The
length of the code is $n$ and for the other parameters, the following notations will be used:
\begin{itemize}
\item $K$ is the size of the code.
\item $k$ is the dimension of the code; since it is not necessarily a linear code its dimension is defined
by:
$$k = \log_Q K \in {\mathbb R}$$
\item $d$ is the minimal distance of the code. For arbitrary non-linear codes, $d$ corresponds to the minimal distance
between any two codewords. However, since $C({\cal P},E)$ is $\F_q$ linear, $d$ is also equal to the minimal
weight of a non-zero codeword.
\end{itemize}
Note that if the evaluation map is injective, then $K = q^{h^0(\C, E)}$ and therefore $k = \displaystyle
\frac{h^0(\C, E)}{r}$.

\begin{theo}
Assume that $E$ is a weakly stable vector bundle of degree $e$ and slope $\mu = e/r < n$. Then:

$(a)$ the evaluation map is injective,

$(b)$ $d\geq n-\lfloor\mu\rfloor$,

$(c)$ $k \geq \mu + 1 - g$,

$(d)$ the Singleton defect of the code is uperbounded by the curve genus $g$.
\end{theo}
{\em Proof}. Since $E$ is weakly stable, any section $f\in H^0(\C, E)$ vanishes in at most $\lfloor\mu\rfloor$ points, which proves $(a)$
and $(b)$. Because the evaluation map is injective $(a)$, we also have\break $k = h^0(\C, E)/r$. By the Riemann-Roch theorem $h^0(\C,
E)\geq e + r(1-g)$, therefore $k \geq \mu + 1-g$. Finally, $(d)$ follows from $(b)$ and $(c)$. \hfill $\square$

\begin{prop}
Let $E$ be vector bundle of degree $e$ and slope $\mu = e/r$. Assume that both $E$ and $E^{*}$ are weakly stable and $\mu > 2g-2$. Then $k
= \mu + 1-g$.
\end{prop}
{\em Proof.} The tensor product $\Omega\otimes E^{*}$ is a weakly stable vector bundle of degree:
$$\deg(\Omega\otimes E^{*}) = r \deg(\Omega) - \deg(E) = r(2g-2) - e < 0$$
Therefore $h^0(\Omega\otimes E^{*}) = 0$ and the assertion follows by the Riemann-Roch theorem. \hfill $\square$

\section{Decoding algorithm} \label{sec:decoding}

 Let $C({\cal P},E)$ be a code over $\F_Q$ defined by a rank-$r$ vector bundle $E$ and an
evaluation set ${\cal P}=\{P_1,\dots,P_n\}$. Assume that the codeword $(f(P_1), f(P_2), \dots, f(P_n))$,
defined by
 some $f\in H^0(\C, E)$, is transmitted over the $Q$-ary symmetric channel and let $(y_1, y_2, \dots,
y_n)$ be the received word. Our goal is to decode the codeword and for this we proceed in a way similar to \cite{interleaved_ag}. Let $t$
be a parameter to be determined latter and let $L$ be a line bundle of degree $l := t + g$. The decoding works in two steps as follows:
\begin{list}{$\bullet$}{\itemindent -1.95\parindent \leftmargin 9mm }
\item[$(S1)$] Find a non-zero element $$(v,w)\in H^0(\C, E\otimes L) \times H^0(\C, L)$$ such that
$$v(P_i) = y_i \otimes w(P_i), \ \forall i=1,\dots, n$$
If $(v,w)$ does not exist, output a decoding error.
\item[$(S2)$] If there exists $\bar{f}\in H^0(\C, E)$ such that
$$v = \bar{f} \otimes w$$
decode $\bar{f}$. Otherwise, output a decoding error.
\end{list}

\begin{prop}\label{existence_w}
Let $\epsilon$ denote the number of errors incurred during transmission. If $\epsilon \leq t$ then there
exists an non-zero element $(v,w)$ satisfying $(S1)$.
\end{prop}
{\em Proof}. Assume that errors occur in points $P_{i_1},\dots, P_{i_\epsilon}$ and let $D_{\mbox{\scriptsize err}} = P_{i_1}+ \dots +
P_{i_\epsilon}$. Using the Riemann-Roch theorem it can be proved that $h^0(\C, L(-D_{\mbox{\scriptsize err}})) > 0$. Therefore, we can
choose a non-zero $w\in H^0(\C, L(-D_{\mbox{\scriptsize err}}))$ and define $v = f\otimes w$. If $P_i$ is not an error point, meaning that
$P_i\not\in\{P_{i_1},\dots, P_{i_\epsilon}\}$, then $y_i = f(P_i)$ and so $v(P_i) = y_i\otimes w(P_i)$. Otherwise, the equality $v(P_i) =
y_i\otimes w(P_i)$ still holds, because $v(P_i) = w(P_i) = 0$ for any $P_i\in\{P_{i_1},\dots, P_{i_\epsilon}\}$. \hfill $\square$

\begin{theo}
Let $\epsilon$ denote the number of errors incurred during transmission. Assume that $E$ is a weakly stable
vector bundle of degree $e$ and slope $\mu = e/r$, such that:
$$\epsilon \leq t \ \mbox{ and }\ \epsilon + t \leq n - \mu -g$$
Then the above decoder outputs the transmitted codeword.
\end{theo}
{\em Proof.} From the above proposition, there exists a non-zero element $(v, w)\in H^0(\C, E\otimes L)
\times H^0(\C, L)$ satisfying $(S1)$, that is:
$$v(P_i) = y_i \otimes w(P_i),\ \forall i=1,\dots,n$$
Assume that errors occur in points ${\cal P} = \{P_{i_1},\dots, P_{i_\epsilon}\}$ and let $D =
\sum_{P_i\not\in{\cal P}} P_i$. For any $P_i\not\in{\cal P}$ we have $y_i = f(P_i)$ and therefore:
$$(v-f\otimes w)(P_i) = y_i \otimes w(P_i) - f(P_i) \otimes w(P_i) = 0$$
It follows that $ v - f\otimes w\in H^0(\C, E\otimes L(-D))$. On the other hand, knowing that $\deg(E) = e$,
$\deg(L) = t+g$ and $\deg(D) = n-\epsilon$, we get:
$$\begin{array}{c@{\,\,}l}
\deg(E\otimes L(-D)) & = e + r(t+g - (n-\epsilon)) \\
&= r(\epsilon + t -n + \mu + g) < 0
\end{array}$$
Consequently, $E\otimes L(-D)$ is a  weakly stable vector bundle of negative degree, so it has no non-zero
global sections. Hence $v = f\otimes w$ and the decoder outputs $f$. \hfill $\square$

Note that the designed correction capacity of the code for the $Q$-ary symmetric channel is $t^{*} = \displaystyle\left\lfloor
\frac{n-\mu}{2} \right\rfloor$. From the above theorem, it follows that the decoding algorithm corrects any pattern of $\epsilon < t^{*} -
\displaystyle\frac{g}{2}$ errors. We note that the above algorithm can easily be extended to correct both errors and erasures.

\medskip
Throughout the rest of this section we give a possible realization the decoding algorithm. Our goal is just to prove that the decoding
algorithm is constructible and executable in polynomial time. We fix once for all:
\begin{itemize}
\item a basis of $\F_Q$ over $\F_q$,
\item a basis of $E_P$ over $\F_q$, for each $P\in{\cal P}$,
\item a basis of $L_P$ over $\F_q$, for each $P\in{\cal P}$,
\item $f_1,\dots,f_h$ a basis of $H^0(\C, E)$ over $\F_q$,
\item $\varphi_1,\dots,\varphi_a$ a basis of $H^0(\C, E\otimes L)$ over $\F_q$,
\item $\psi_1,\dots,\psi_b$ a basis of $H^0(\C, L)$ over $\F_q$
\end{itemize}
The first three basis allow us to identify:
$$E_P\otimes L_P \simeq E_P \simeq \F_Q \simeq \F_q^r$$
Let $(y_1, y_2, \dots, y_n)$ be the received word. For each $1\leq i\leq n$, we consider that
$$y_i \in\F_Q \simeq \left(\begin{array}{c} y_{i,1} \\ \vdots \\ y_{i,r} \end{array}\right)\in \F_q^r$$
and we set
 $$Y = \left(\begin{array}{cccc}
      y_1 &      &        &     \\
          & y_2  &        &     \\
          &      & \ddots &     \\
          &      &        & y_n
      \end{array}\right) \in M_{nr,n}(\F_q)$$
where each $y_i$ is identified with the corresponding column vector. Moreover, we define:
$$FP_i = \left(f_1(P_i),f_2(P_i),\dots,f_h(P_i)\right) \in M_{1,h}(\F_Q)\simeq M_{r,h}(\F_q)$$
$$F = \left(\begin{array}{cccc}
      FP_1 &      &        &     \\
          & FP_2  &        &     \\
          &      & \ddots &     \\
          &      &        & FP_n
      \end{array}\right)
      \begin{array}{c}
        \raisebox{-8mm}{$\in  M_{n,nh}(\F_Q)$} \\
        \raisebox{2mm}{\rotatebox{270}{$\simeq$}}\\
        \raisebox{5mm}{\ \ $M_{nr,nh}(\F_q)$}
    \end{array}$$
$$V = \left(\begin{array}{cccc}
    \varphi_1(P_1) & \varphi_2(P_1) & \cdots & \varphi_a(P_1) \\
    \varphi_1(P_2) & \varphi_2(P_2) & \cdots & \varphi_a(P_2) \\
    \vdots         & \vdots         & \ddots & \vdots         \\
    \varphi_1(P_n) & \varphi_2(P_n) & \cdots & \varphi_a(P_n)
    \end{array}\right)
    \begin{array}{c}
        \raisebox{-8mm}{$\in  M_{n,a}(\F_Q)$} \\
        \raisebox{2mm}{\rotatebox{270}{$\simeq$}}\\
        \raisebox{5mm}{\ \ $M_{nr,a}(\F_q)$}
    \end{array}$$
where all $f_j(P_i)$ and $\varphi_j(P_i)$ are identified with $r\times 1$ column vectors in $\F_q^r$, and
$$W = \left(\begin{array}{cccc}
    \psi_1(P_1) & \psi_2(P_1) & \cdots & \psi_b(P_1) \\
    \psi_1(P_2) & \psi_2(P_2) & \cdots & \psi_b(P_2) \\
    \vdots      & \vdots      & \ddots & \vdots         \\
    \psi_1(P_n) & \psi_2(P_n) & \cdots & \psi_b(P_n)
    \end{array}\right) \in M_{n,b}(\F_q)$$

The decoding algorithm can now be described as follows:
\begin{list}{$\bullet$}{\itemindent -1.95\parindent \leftmargin 9mm }
\item[$(S1)$] Find a non-zero solution $$(v_1,\dots,v_a,w_1,\dots,w_b)\in\F_q^{a+b}$$ of the system
$$V\left(\begin{array}{c}v_1 \\ \vdots \\ v_a \end{array}\right) =
 YW\left(\begin{array}{c}w_1 \\ \vdots \\ w_b \end{array}\right)$$
If the system does not have a non-zero solution, output a decoding error.
\item[$(S2)$] Find a solution $\lambda = (\lambda_1,\dots,\lambda_h)\in\F_q^h$ of the system
 $$V\left(\begin{array}{c}v_1 \\ \vdots \\ v_a \end{array}\right) =
 F\Lambda W\left(\begin{array}{c}w_1 \\ \vdots \\ w_b \end{array}\right)$$
where
$$\Lambda = \left(\begin{array}{ccc}
                ^t\lambda &        &           \\
                          & \ddots &           \\
                          &        & ^t\lambda
                \end{array}\right) \in M_{nh,n}(\F_q)$$
and output $\bar{f} = \lambda_1 f_1 + \cdots + \lambda_h f_h$. If the system does not have any solution, output a decoding error.
\end{list}
Note that the second step of the above realization compute a section $\bar{f} = \lambda_1 f_1 + \cdots + \lambda_h f_h$ verifying
 $$v(P_i) = \bar{f}(P_i)\otimes w(P_i),\ \forall i = 1,\dots,n$$ where $v = v_1\varphi_1+\cdots+v_a\varphi_a$ and $w =
w_1\psi_1+\cdots+w_b\psi_b$. This is a little bit different from the second step of the decoding algorithm, which requires the above
equality to hold for any point $P$ ({\em i.e.} $v = \bar{f}\otimes w$). Assuming that the vector bundle $E$ is weakly stable, any non-zero
section of $E\otimes L$ vanishes in at most $\mu(E\otimes L) = \mu+t+g$ points. Furthermore, assuming that $n > \mu+t+g$ and $v(P_i) =
\bar{f}(P_i)\otimes w(P_i),\ \forall i = 1,\dots,n$, it follows that the section $v-\bar{f}\otimes w$ vanishes in more than $\mu+t+g$
points, and so $v=\bar{f}\otimes w$.

\section{Construction of weakly stable vector bundles}\label{sec:construction_vb}
Most of statements concerning the parameters and the decoding of algebraic codes constructed from vector bundles require weakly stable
vector bundles. A trivial example of a weakly stable vector bundle of rank $r$ and degree $e$ is given by the direct sum of $r$ line
bundles of degree $e$. Such a vector bundle is called completely undecomposable. In this section we show that for any curve of genus $g\geq
2$ and any integers $r>0$ and $e$ there exist non-trivial examples of weakly stable vector bundles of rank $r$ and degree $e$.

Let $e = \alpha r +\beta$, with $\alpha, \beta \in {\mathbb Z}$ such that $0\leq \beta < r$. Let $F_1, F_2$
and $F$ be line bundles on $\C$, such that:
 $$\deg(F_1) = \deg(F_2) = \alpha, \ \ \deg(F) = \alpha+1$$
Consider a sequence of vector bundles $E_i$ defined by the following non-trivial extensions:
$$\begin{array}{c@{\,\,}c@{\,\,}c@{\,\,}c@{\,\,}c@{\,\,}c@{\,\,}c}
\multicolumn{7}{c}{E_1 = F_1}\\
 0 \rightarrow & E_1           & \longrightarrow & E_2           & \longrightarrow & F_2 & \rightarrow 0\\
\multicolumn{7}{c}{\cdots\cdots\cdots\cdots\cdots\cdots\cdots\cdots\cdots\cdots\cdots\cdots}\\
 0 \rightarrow & E_{r-\beta-1} & \longrightarrow & E_{r-\beta}   & \longrightarrow & F_2 & \rightarrow 0\\
 0 \rightarrow & E_{r-\beta}   & \longrightarrow & E_{r-\beta+1} & \longrightarrow & F   & \rightarrow 0\\
 \multicolumn{7}{c}{\cdots\cdots\cdots\cdots\cdots\cdots\cdots\cdots\cdots\cdots\cdots\cdots}\\
  0 \rightarrow & E_{r-1}      & \longrightarrow & E_{r}         & \longrightarrow & F   & \rightarrow 0
 \end{array}$$
The extensions of $F_2$ by $E_i$, $i\leq r-\beta-1$, are classified by $H^1(\C, F^{*}_2\otimes E_i) \simeq H^0(\C, \Omega\otimes F_2\otimes
E_i^{*})$, by Poincar\'e duality. Since $F^{*}_2\otimes E_i$ is a vector bundle of degree $0$ and rank $i$, by the Riemann-Roch theorem we
obtain $h^0(\C, \Omega\otimes F_2\otimes E_i^{*}) > 0$. It follows that there exist non-trivial extensions of $F_2$ by $E_i$. Using similar
arguments, it can also be shown that there exist non-trivial extensions of $F$ by $E_i$, $i\geq r-\beta$.
\begin{prop}
Let $E := E_r$. Then:
\begin{itemize}
\item[$(i)$] $\deg(E) = \alpha r + \beta = e$ and $\mbox{rk}(E) = r$.
\item[$(ii)$] $E$ is weakly-stable.
\end{itemize}
\end{prop}
{\em Proof.} $(i)$ is clear from construction. To see that $E$ is weakly stable, consider $L$ a line sub-bundle of $E$ and let $i$ be the
smallest index such that $L$ is contained in $E_i$. Then the morphism $L\rightarrow E_i \rightarrow E_i/E_{i-1}$ is non zero, therefore:
$$\deg(L) \leq \deg (E_i/E_{i-1}) \leq \alpha+1$$
If $\deg(L) = \alpha+1$ then $L\simeq E_i/E_{i-1}$ and the extension of $E_i/E_{i-1}$ by $E_i$ would split. So $\deg(L) \leq \alpha \leq
\mu(E)$, which proves that $E$ is weakly stable. \hfill $\square$

\section{Conclusions}\label{sec:conclusions}
A new construction of AG codes from vector bundles on algebraic curves was proposed in this paper, which allows a unified treatment of
classical AG codes and more recently interleaved AG codes. In the same time, this construction extends the above class of AG codes to a
much larger class of codes. These new codes have very good properties and they can be designed over very large Galois fields with
reasonable decoding complexity, since decoding can be performed in a smaller field. We have also provided a decoding algorithm for these
codes that corrects any number of errors up to $t^{*} - g/2$, where $t^{*}$ is the designed correction capacity of the code and $g$ is the
curve genus.

The aim of this paper is also to relate the construction of AG codes to more sophisticated and powerful concepts in algebraic-geometry.
However, this is only a first step and more work has to be done in this area. It is very likely that for suitable choices of vector bundles
$E$ and $L$, the decoding algorithm will correct errors up to the designed correction capacity of the code. We think that future work could
bring out many  useful interactions between algebraic geometric codes and vector bundles on algebraic curves.

\bibliographystyle{abbrv}
\vspace{-1mm}
\bibliography{./bib/MyBiblio,./bib/Zotero}

\begin{thebibliography}{1}

\bibitem{ag_codes_survey}
I.~Blake, C.~Heegard, T.~H{\o}holdt, and V.~Wei.
\newblock Algebraic-geometry codes.
\newblock {\em IEEE Trans. Inform. Theory}, 44(6):2596--2618, 1998.

\bibitem{interleaved_rs}
D.~Bleichenbacher, A.~Kiayias, and M.~Yung.
\newblock Decoding interleaved {R}eed {S}olomon codes over noisy channels.
\newblock In {\em ICALP 2003, Proceedings of}, pages 97--108, 2003.

\bibitem{interleaved_ag}
A.~Brown, L.~Minder, and M.~A. Shokrollahi.
\newblock Improved decoding of interleaved {AG} codes.
\newblock {\em Lecture Notes in Computer Science}, 3796:37--46, 2005.

\bibitem{goppa_curves}
V.~D. Goppa.
\newblock Codes on algebraic curves.
\newblock {\em Dokl. Acad. Nauk SSSR}, 259:1289--1290, 1981.
\newblock Translation: Soviet Math. Dokl., vol. 24, pp. 170-172, 1981.

\bibitem{goppa_ag_codes}
V.~D. Goppa.
\newblock Algebraic-geometric codes.
\newblock {\em Izv. Akad. Nauk SSSR}, 46, 1982.
\newblock Translation: Math. USSR Izv., vol. 21, pp. 75-91, 1983.

\bibitem{hartshorne}
R.~Hartshorne.
\newblock {\em Algebraic geometry}.
\newblock Graduate Texts in Mathematics. Springer-Verlag, 1977.

\bibitem{decoding_ag_survey}
T.~H{\o}holdt and R.~Pellikaan.
\newblock On the decoding of algebraic-geometric codes.
\newblock {\em IEEE Trans. Inform. Theory}, 41(6):1589--1614, 1995.

\end{thebibliography}

\end{document}